%% file: geomet20Dec21.tex
\documentclass[review]{elsarticle}
\usepackage{longtable}
\usepackage{lscape}

\usepackage{graphicx}
\usepackage{amssymb}

\def\t{{^T}}
\def\dx{{\partial\over\partial x_1}}
\def\dy{{\partial\over\partial x_2}}
\def\dz{{\partial\over\partial x_3}}

\def\dfx{{\partial f\over\partial x_1}}
\def\dfy{{\partial f\over\partial x_2}}

\def\R{{\mathbf{R}}}

\def\<{{\langle}}
\def\>{{\rangle}}
\begin{document} % for ELSEVIER

\begin{frontmatter} % for ELSEVIER

\title{A geometric analysis of nonlinear dynamics and its application to financial time series}

% for ELSEVIER
\author[tus]{Isao Shoji\corref{shoji}}
\ead{shoji@rs.tus.ac.jp}
\address[tus]{Tokyo University of Science,
Fujimi, Chiyoda, Tokyo 102-0071, Japan}
\author[tus]{Masahiro Nozawa}

\cortext[shoji]{Corresponding author}

\begin{abstract}
A geometric method to analyze nonlinear oscillations is discussed.
We consider a nonlinear oscillation modeled by a second order ordinary differential equation without specifying the function form.  By transforming the differential equation into the system of first order ordinary differential equations, the trajectory is embedded in $R^3$ as a curve, and thereby the time evolution of the original state can be translated into the behavior of the curve in $R^3$, or the vector field along the curve. We analyze the vector field to investigate the dynamic properties of a nonlinear oscillation.
While the function form of the model is unspecified, the vector fields and those associated quantities can be estimated by a nonparametric filtering method.
We apply the proposed analysis to the time series of the Japanese stock price index.
The application shows that the vector field and its derivative will be used as the tools of picking up various signals that help understanding of the dynamic properties of the stock price index.
\end{abstract}

\begin{keyword} % for ELSEVIER
Stochastic differential equation;
Nonparametric filter; Vector field; Geodesics;
\end{keyword} % for ELSEVIER

\end{frontmatter} % for ELSEVIER

%\setlength{\baselineskip}{2pc}
%\setlength{\baselineskip}{1.5pc}
% main text
%\newpage
\section{Introduction}
Nonlinear oscillations are ubiquitous phenomena from natural science to social science. Various method of analyzing their dynamic properties have been proposed. For example, 
neural oscillations are modeled by nonlinear oscillations which are estimated from EEG time series (Riera et al., 2004, Stephan, et al., 2008, Valdes-Sosa et al., 2009, Havlicek et al., 2011, Livina et al., 2011),
heart rate variability is modeled by nonlinear oscillations to monitor the states of the heart (Grudzinski and Zebrowski, 2004, Gois and Savi 2009, Zebrowski et al 2007), 
population dynamics in epidemiology are analyzed by stochastic differential equations (Gao et al, 2019, Wang et al, 2018), 
the light curves of active galactic nuclei in astronomy  (Misra and Zdziarski, 2008, Phillipson et al., 2018), and dynamic behavior of asset prices in finance (Chiarella et al., 2009, Date and Ponomareva, 2011), and so on.

Mathematically a nonlinear oscillation can be modeled by the following second order ordinary differential equation:
%\[
\begin{equation}\label{df:org}
\ddot x=f(x,\dot x).
\end{equation}
%\]
This differential equation is equivalent to the system of the first order ordinary differential equations:
\begin{eqnarray*}\label{df:sys}
dx_1&=&x_2dt\\
dx_2&=&f(x_1,x_2)dt
\end{eqnarray*}
where $x_1=x$ and $x_2=\dot x$.
Since the dynamic behavior of the solution, or trajectory, $x(t)$ corresponds to that of $(x_1(t),x_2(t))$ in $R^2$, the latter
is more useful to analyze the dynamic behavior from a geometric point of view, and thus we focus on the trajectory of the first order differential equation instead of the original one (\ref{df:org}).
Moreover, the trajectory of $(x_1(t),x_2(t))$ is more informative because it contains the information on the velocity.
Hence, more merit will be expected in analyzing the trajectory of $(x_1(t),x_2(t))$ than that of $x(t)$.

Here, we should note that the behavior of $(x_1(t),x_2(t))$ is restricted by $f$ in (\ref{df:sys}). To express the restriction by $f$ more explicitly, we may embed the trajectory of $(x_1(t),x_2(t))$ into the 3-dimensional Euclidean space $R^3$ by constructing a manifold $M$ by $M=\{(x_1,x_2,x_3)| x_3=f(x_1,x_2)\}$. Then, the behavior of $(x_1(t),x_2(t))$ is transformed into that of $(x_1(t),x_2(t),x_3(t))$ where $x_3(t)=f(x_1(t),x_2(t))$. The behavior of $(x_1(t),x_2(t),x_3(t))$ is more informative because it contains the information on the acceleration.
Hence, we will enjoy such a merit as discussed in this paper, by analyzing the trajectory, or more geometrically speaking the curve $(x_1(t),x_2(t),x_3(t))$ in $R^3$ instead of the curve $(x_1(t),x_2(t))$ in $R^2$.

The dynamics of $\gamma(t)=(x_1(t),x_2(t),x_3(t))$ can be characterized by the vector field $V=d\gamma/dt$ along the curve $\gamma(t)$; in other words, to know $V$ is to know the dynamics of $\gamma(t)$.
Moreover, the covariant derivative of $V$ along $\gamma(t)$, denote $\nabla V$,  also conveys useful information on the curve $\gamma(t)$. For example, if $\nabla V=0$, $\gamma(t)$ is a geodesic, implying that among the curves joining $p$ to $q$ $(p,q\in M)$ the length of $\gamma$ from $p=\gamma(t_1)$ to $q=\gamma(t_2)$ is the shortest. When considering the transition from $p$ to $q$, if the covariant derivative $\nabla V=0$, the transition by $\gamma$ is the most efficient in a certain sense. Hence, by checking whether $\nabla V=0$ from moment to moment, we can decide whether or not the transition implied by an oscillation is efficient.

The function form of $f$ plays a crucial role in the dynamic behavior of $\gamma(t)$. For real applications, however, it is quite difficult to specify the function form beforehand. 
Rather, the information on the function form needs to be derived from data. To this end, nonparametric methods may be useful. However, every nonparametric method is not necessarily applicable to our setting because in general we can not expect to observe all the data necessary for the analysis above; $x_1$ could be observed but, generally speaking, neither $x_2$ nor $x_3$ is expected to be observed.
In fact, since nonparametric modeling based on regression (Fan and Gijbels, 1998, Fan and Yao, 2005, Fan and Zhang, 2003) requires an objective variable as well as explanatory variables, the modeling does not work under the circumstances that only partially observed data is available. To this end, methods of filtering may be useful.
Hence, the nonparametric filtering method discussed in Shoji (2020) is used in this paper. Even from partially observed data, the method enables us to estimate unobservable states by the method of filtering. 

By applying the proposed analysis together with the nonparametric filtering method, we conduct an empirical analysis of financial time series. Using daily data of the Japanese stock price index, or Nikkei225, of sample size more than 10,000, we investigate the dynamic properties of the index price by the vector field and its covariant derivative. With estimated vector fields and their covariant derivatives we compare the transitive properties of Nikkei225 from year to year through relating those behaviors with the historical events.

The paper is organized as follows.
In section 2 we explain the geometric method for analyzing the time series of the light ratio.
In section 3 we develop a nonparametric method to estimate quantities used for the above analysis.
An application of the proposed analysis to the financial time series are explained in section 4.
We conclude in the final section.

%%%%%%%%%%%%%%%%%%%%%%%%%%%%%%%%%%%%%%%%%%%%%%%%%
\section{Geometric analysis}
As in Shoji et al (2020), we suppose that a nonlinear oscillation satisfies the following 2nd order differential equation:
\begin{eqnarray}
\ddot x-f(x,\dot x)&=&0.\label{ode:org}
\end{eqnarray}
This differential equation can be reformulated by,
\begin{eqnarray}
dx_{1}&=&x_{2}dt,\label{ode:x1}\\
dx_{2}&=&f(x_{1},x_{2})dt,\label{ode:x2}
\end{eqnarray}
where $x_1= x$, $x_2=dx_1/dt$.

Usually the curve of $(x_1(t),x_2(t))$ is analyzed in the phase space with $x_1$ and $x_2$ axis. However, 
since the dynamics of $(x_1(t),x_2(t))$ is affected by $f$ in (\ref{ode:x2}), the influence of $f$ on the dynamics should be taken into account. Thus, we embed the curve of $(x_1(t),x_2(t))$ into the 3-dimensional Euclidean space $R^3$ by constructing a manifold $M$ by $M=\{(x_1,x_2,x_3)\in R^3| x_3=f(x_1,x_2)\}$. Here note that three coordinates $x_1$ to $x_3$ in $M$ mean the original state, its velocity, and its acceleration.
This coordinate system is well known to the so-called differential phase space  and is convenient to grasp the dynamic behavior of the curve geometrically. %, and used in previous 
Let $\gamma(t)$ be the point of $M$ at time $t$. Given an interval $I\subset R$, for $t\in I$ $\gamma(t)$ shows the curve on $M$, more specifically it moves with satisfying the system of differential equations because $x_2=\dot x_1$ and $x_3=\dot x_2$.
$\gamma(t)$ is considered to be more informative than the original state $x_1(t)$ itself because $\gamma(t)$ additionally contains the information on velocity and acceleration. Hence, we are mainly interested in $\gamma(t)$. In particular, the dynamic behavior of $\gamma(t)$ can be characterized by the vector field along $\gamma(t)$, or $V=d\gamma/dt$, which belongs to the tangent space $T_{\gamma(t)}(M)$.
While $V$ determines the local behavior of $\gamma(t)$, the integral curve of $V$ becomes $\gamma(t)$ and thus we can grasp the global behavior of $\gamma(t)$ through $V$ as well.

\subsection{Covariant derivative}
The time change of $\gamma(t)$ is characterized by $V$, and thus it sounds reasonable that the time change of $V$ can be characterized by the derivative of $V$, or $dV/dt$. However, this is not necessarily true because $\gamma(t)$ lies in $M$ so that the component of $dV/dt$ perpendicular to $T_{\gamma(t)}M$ will have no effect on the behavior of $V$.
Hence, the projection of $dV/dt$ to $T_{\gamma(t)}M$ is needed, which is given by the covariant derivative of $V$ along $\gamma(t)$, denoted by  $\nabla V$ in this paper.

Since $\nabla V$ is a kind of derivatives of $V$, $\nabla V$ will give useful information on the dynamics of $V$. For example, if the derivative of $V$ is zero, it may be reasonable to consider that $\|V\|$ is constant. This intuition is correct if $\gamma(t)$ is a geodesic.
Actually, because $\<V,V\>$ is scalar, $\nabla\<V,V\>={d\over dt}\<V,V\>$. 
On the other hand, because $\nabla$ is compatible with the metric $\<\cdot,\cdot\>$ and the metric is symmetric, $\nabla\<V,V\>=2\<\nabla V,V\>$. Because $\nabla V=0$ if $\gamma(t)$ is a geodesic, ${d\over dt}\<V,V\>=0$, implying $\|V\|$ is constant. 

Moreover, if $\gamma(t)$ is a geodesic joining $p$ to $q$ in $M$, the length of $\gamma(t)$ from $p$ to $q$ is the shortest among those of curves joining $p$ to $q$; a geodesic gives the shortest path form $p$ to $q$. This implies that $\gamma(t)$ will produce an efficient behavior in a certain sense if it is a geodesic.  Hence, it may be interesting to see whether $\gamma(t)$ is a geodesic, which can be easily seen by checking whether $\nabla V=0$.

$\nabla V$ can be expressed as the linear combination of the basis $E_1,E_2$ of $T_{\gamma(t)}M$, $\beta_1E_1+\beta_2E_2$, whose coefficient vector $\beta=(\beta_1,\beta_2)\t$ is given by,
\begin{eqnarray*}
\beta&=&(X\t X)^{-1}X\t {dV\over dt},\\
X&=&
\left(
\begin{array}{cc}
1& 0\\
0& 1\\
{\partial f\over\partial x_1}& {\partial f\over\partial x_2}
\end{array}
\right).
\end{eqnarray*}
with,
\begin{eqnarray}
dV/dt&=&\dot x_2\dx+\dot f\dy+\ddot f\dz,\label{eq:dV/dt}\\
&=&f\dx+({\partial f\over\partial x_1}x_2+{\partial f\over\partial x_2}f)\dy\nonumber\\
&&+({\partial^2 f\over\partial x_1^2}x_2^2+2{\partial^2 f\over\partial x_1\partial x_2}x_2f+{\partial f\over\partial x_1}f\nonumber\\
&&+
{\partial^2 f\over\partial x_2^2}f^2+{\partial f\over\partial x_2}({\partial f\over\partial x_1}x_2+{\partial f\over\partial x_2}f))\dz,\\
E_1&=&\dx+\dfx\dz,\label{base1}\\
E_2&=&\dy+\dfy\dz.\label{base2}
\end{eqnarray}
See Shoji et al (2020) for the detail derivation.

%%%%%%%%%%%%%%%%%%%%%%%%%%%%%
\section{Estimation method}\label{sec:npf}
The geometric analysis explained in the previous section can be carried out if $f$ is known. Actually, however, $f$ is unknown in almost all the real applications. Besides, we have almost no way to retrieve necessary information on its function before analysis.
Furthermore, we are usually unable to access all the information on $\gamma(t)$; it is often the case that $x_1$ is observable but neither $x_2$ nor $x_3$. So, we assume only $x_1$ to be observable in the following.
Under these restrictions, we try to estimate necessary quantities required for the analysis.

Let $X_{1,t}$ and $X_{2,t}$ be stochastic processes corresponding to $x_1$ and $x_2$ in (\ref{ode:x1})-(\ref{ode:x2}), respectively.  
The processes are assumed to follow the SDE below:
\begin{eqnarray}
dX_{1,t}&=&X_{2,t}dt+\sigma_1 dB_{1,t},\label{sde:X1}\\
dX_{2,t}&=&f(X_{1,t},X_{2,t})dt+\sigma_2 dB_{2,t},\label{sde:X2}
\end{eqnarray}
where $(B_{1,t},B_{2,t})$ is a 2-variate standard Brownian motion with constant diffusion coefficients $\sigma_1$ and $\sigma_2$.
Here, no function form of $f$ is assumed.
In addition, only $X_{1,t}$ is observable and it contains observation error.
So, given observed time series of $X_{1,t}$, denote $\{Z_{t_k}\}_{1\le k\le n}$ $(\Delta t=t_k-t_{k-1})$, $Z_{t_k}$ is associate with $X_{1,t}$ by,
\begin{eqnarray}
Z_{t_k}&=&X_{1,t_k}+\varepsilon_k,\label{obseq}
\end{eqnarray}
where $\{\varepsilon_k\}_{1\le k\le n}$ is assumed to be independently identically normally distributed with mean zero and variance $\sigma_\varepsilon^2$.

Under the settings above, to estimate the SDE without specifying $f$, a nonparametric method may be required.
In addition, the model has to be estimated from partially observed data.
To meet these requirements, we apply a nonparametric filtering (NPF) method discussed in Shoji (2020). 
According to the NPF method of the third order expansion, the discretized version of the state space model is given as follows.
Let the state vector $\xi_k$ at discrete time $t_k$ be a vector in $\R^8$, which is given by,
\begin{equation}
\xi_k=(X_{1,t_k},X_{2,t_k},Y^{(0,0)}_{t_k},Y^{(1,0)}_{t_k},Y^{(0,1)}_{t_k},Y^{(2,0)}_{t_k},Y^{(1,1)}_{t_k},Y^{(0,2)}_{t_k})\t ,
\end{equation}
where,
\begin{eqnarray*}
Y^{(0,0)}_{t}&=&f(X_{t}),\ Y^{(1,0)}_{t}={\partial f\over\partial x_1}(X_{t}),\ Y^{(0,1)}_{t}={\partial f\over\partial x_2}(X_{t}),\\
Y^{(2,0)}_{t}&=&{\partial^2 f\over\partial x_1^2}(X_{t}),\ Y^{(1,1)}_{t}={\partial^2 f\over\partial x_1\partial x_2}(X_{t}),\ Y^{(0,2)}_{t}={\partial^2 f\over\partial x_2^2}(X_{t}).
\end{eqnarray*}
Because we have no assumption about the function form of $f$, we also need to estimate the values of the function and its derivatives from  $\{Z_{t_k}\}_{1\le k\le n}$. This is why additional six states are considered as unobservable states.

The state vector $\xi_k$ satisfies the following system and observation equations. See appendix for the detail formula. 
\begin{eqnarray}
\xi_{k+1}&=&F_k\xi_k+c_k+e_{k+1},\\
Y_{t_k}&=&H\xi_k+\varepsilon_{t_k}.
\end{eqnarray}
Applying the Kalman updating formula to the above state space model, the filter and prediction of $\xi_k$, which are given as $\xi_{k|k}=E[\xi_k|\{Z_{t_j}\}_{1\le j\le k}] $ and $\xi_{k+1|k}=E[\xi_{k+1}|\{Z_{t_j}\}_{1\le j\le k}] $, can be computed as follows.
\begin{eqnarray}
\xi_{k+1|k}&=&F_k\xi_{k|k}+c_k,\\
\Sigma_{k+1|k}&=&F_k\Sigma_{k|k}F_k^T+Q_k,\\
K_k&=&\Sigma_{k|k-1}H^T(H\Sigma_{k|k-1}H^T+\sigma_\varepsilon^2)^{-1},\\
\xi_{k|k}&=&\xi_{k|k-1}+K_k(X_{t_k}-H\xi_{k|k-1}),\\
\Sigma_{k|k}&=&(I-K_kH)\Sigma_{k|k-1},
\end{eqnarray}
where,
\begin{eqnarray*}
\Sigma_{k+1|k}&=&
E\left[(\xi_{k+1}-\xi_{k+1|k})(\xi_{k+1}-\xi_{k+1|k})^T
\left|\{Z_{t_j}\}_{1\le j\le k}\right.\right],\\
\Sigma_{k|k}&=&
E\left[(\xi_k-\xi_{k|k})(\xi_k-\xi_{k|k})^T
\left|\{Z_{t_j}\}_{1\le j\le k}\right.\right],\\
Q_k&=&
E\left[e_{k+1}e_{k+1}^T\left|\{Z_{t_j}\}_{1\le j\le k}\right.\right].
\end{eqnarray*}
Then the estimates of the unobservable states of our interest can be found in the components of $\xi_{k|k}$ or $\xi_{k+1|k}$. For example, the estimate of the unobservable state $X_{2,t_k}$ is given by the second component of $\xi_{k|k}$, and the predicted values of $X_{1,t_{k+1}}$ and $X_{2,t_{k+1}}$ are given by the first and second components of $\xi_{k+1|k}$ and so on.

Here note that the system has several parameters to be estimated; the coefficients of deviation $\sigma_1$, $\sigma_2$, and $\sigma_\varepsilon$ and the nuisance parameters $\theta_0,\theta_1,\theta_2$, and $\theta_3$, which correspond to $Y^{(3,0)}_t,Y^{(2,1)}_t,Y^{(1,2)}_t$, and $Y^{(0,3)}_t$. These parameters can be estimated by the quasi-maximum likelihood estimation from $\{Z_{t_k}\}_{1\le k\le n}$ by using the following likelihood function.  For the parameter vector $\theta=(\sigma_1,\sigma_2,\sigma_\varepsilon,\theta_0,\theta_1,\theta_2,\theta_3)$,
\begin{eqnarray}
L(\theta)&=&\Pi_{k=1}^{n-1}p(Z_{t_{k+1}}|Z_{t_{k}};\theta),\label{ml:L}\\
p(Z_{t_{k+1}}|Z_{t_k};\theta)&=&
(2\pi (H\Sigma_{k+1|k}H^T+\sigma_\varepsilon^2))^{-1/2}\label{ml:p}\\
&&\times\exp\left\{-{(Z_{t_{k+1}}-H\xi_{k+1|k})^2\over 2(H\Sigma_{k+1|k}H^T+\sigma_\varepsilon^2)}\right\}\nonumber.
\end{eqnarray}
Then maximizing $L(\theta)$ with respect to $\theta$, the maximum likelihood estimate $\hat\theta$ is obtained.

%%%%%%%%%%%%%%%%%%%%%%%%%%%%%
\subsection{Estimation of vector fields}
For $\gamma(t)=(x_1(t),x_2(t),x_3(t))\t$, $V=dV/dt$ and thus $\|V\|$ is given  by,
\begin{eqnarray}
\|V\|&=&\sqrt{\dot x_1^2+\dot x_2^2+\dot x_3^2},\label{eq:|V|}\\
\dot x_1&=&x_2,\nonumber\\
\dot x_2&=&f(x_1,x_2),\nonumber\\
\dot x_3&=&{\partial f\over\partial x_1}\dot x_1+{\partial f\over\partial x_2}\dot x_2.\nonumber
\end{eqnarray}
Here note that $x_1,x_2,x_3$ are deterministic variables.
Thus they need to be replaced by their estimates obtained from the NPF method.
Application of the NPF method produces $\xi_{k|k}$, thereby substituting the values of the variables for the components of $\xi_{k|k}$ as follows:
\begin{eqnarray}
x_1(t_k)&=&X_{1,t_k|t_k},\ x_2(t_k)=X_{2,t_k|t_k},\\
f(t_k)&=&Y^{(0,0)}_{t_k|t_k},\\
{\partial f\over\partial x_1}(t_k)&=&Y^{(1,0)}_{t_k|t_k},\ {\partial f\over\partial x_2}(t_k)=Y^{(0,1)}_{t_k|t_k},\\
{\partial^2 f\over\partial x_1^2}(t_k)&=&Y^{(2,0)}_{t_k|t_k},\ {\partial^2 f\over\partial x_1\partial x_2}(t_k)=Y^{(1,1)}_{t_k|t_k},\\
{\partial^2 f\over\partial x_2^2}(t_k)&=&Y^{(0,2)}_{t_k|t_k}.
\end{eqnarray}
Plugging these estimates in the formula of $\|V\|$, we get the estimate of $\|V\|$ at time $t_k$.
As for the covariant derivative, because $\nabla V=\beta_1E_1+\beta_2E_2$ by using $E_1$ and $E_2$ in (\ref{base1}) and (\ref{base2}), the estimates of $\nabla V$ and its length are obtained in the same way as above. 

%%%%%%%%%%%%%%%%%%%%%%%%%%%%%
\section{Application to financial time series}
We apply the proposed analysis to time series of the Japanese stock price index, called Nikkei225, sampled from 1965 to 2019 on daily basis, whose sample size is more than 10,000 in total.
In the following empirical analysis we use the time series transformed into the logarithm of relative price to the initial one.
We have interest in what information $V$ and $\nabla V$ convey on the dynamic behavior of the stock price index.

From a geometric viewpoint, the larger $\|V\|$ implies the speed of the transition of  $\gamma(t)$ is faster; drastic behaviors of the stock price index may be seen while $\gamma(t)$ itself is not the stock price index but a curve in $R^3$. On the other hand, if $\nabla V=0$, or equivalently $\|\nabla V\|=0$, $\gamma(t)$ is a geodesic, implying that $\gamma(t)$ produces the shortest path among curves joining $p$ to $q$ in the manifold $M$. Thus, the transition of $\gamma(t)$ is thought to be efficient in a certain sense if $\gamma(t)$ is a geodesic; otherwise the path from $p$ to $q$ becomes a detour and the transition is thought to be inefficient.

Taking this geometric interpretation into account, we consider the dynamic behavior of Nikkei225  through those of  $\|V\|$ and  $\|\nabla V\|$.
In the first place, we estimate the parameters of the state space model by applying the nonparametric filtering method of 3rd order expansion to the time series and the result is presented in table \ref{tbl:param}.
Then, using the estimates, we compute the length of $V(t_k)$ and $\nabla V(t_k)$ at time $t_k$, whose time series plots from 1967 to 2019 are displayed in figure \ref{fig1} to \ref{fig9}. Here, the first two years are omitted to remove the influence of the initial estimates in the Kalman filtering. For each figure, the leftmost panel displays the time series plots of the logarithm of the relative price, the center panel for those of $\|V\|$, and the rightmost panel for those of $\|\nabla V\|$.

As indicated in the previous section, $\nabla V=0$ implies $\|V\|$ is constant. Hence, expectedly small $\|\nabla V\|$ implies $\|V\|$ has small change. Actually, looking at the panel of 1991, $\|V\|$ moves around 0.5 from 25d to 100d, where $\|\nabla V\|$ shows small values. 
Because, when $\nabla =0$, $\gamma(t)$ becomes a geodesic, the transition of the index price is considered to be efficient in a certain sense. 

%
% In the following analyis 0.5*(F*F')+G*V*G' is used in mk_vf
%
Although we try to see the behaviors of $\|V\|$ and $\|\nabla V\|$ in more detail, their time series plots show considerably various patterns from year to year. Hence, we conduct an analysis of variance (ANOVA) for the times series of $\|V\|$ and $\|\nabla V\|$. Looking at the group of the highest value of $\|V\|$, or grouping the years in which the sample mean of $\|V\|$ is not significantly different from the highest, it consists of the years $\{$2001,2008$\}$ and the second highest are  and $\{$1972,1987,1989,1992,2013$\}$, respectively. Recalling that $V$ stands for the velocity of $\gamma(t)$, it is expected that steep upward or downward sloping of the index price may occur in the years of the group. In fact, looking at the time series plots, we had such price changes. Besides, it might be interesting to see that 2008 belongs to the group, which is related to the year of the Lehman's collapse. 
On the other hand, focusing on the group of the lowest $\|V\|$, it consists of $\{$1994,2019$\}$. According to their time series plots, so steep upward or downward sloping in the price index is not observed.

Turning our attention to the result of ANOVA for $\|\nabla V\|$, the group of the highest $\|\nabla V\|$ consists of $\{$2001,2008$\}$ and the groups of the second and third highest consist of $\{$1972,2013$\}$ and $\{$1974,1987,1992,2002,2009$\}$, respectively. Because $\gamma(t)$ is thought to be far away from a geodesic, the behaviors of the index price in the years of these groups are considered to be inefficient. Interestingly, 2008 also belongs to these groups, implying the transition of the index price may be inefficient through the Lehman's collapse.

\begin{figure}
\caption{$\|V\|$ and $\|\nabla\|$}
\label{fig:prin}
\includegraphics[height=16cm,width=16cm]{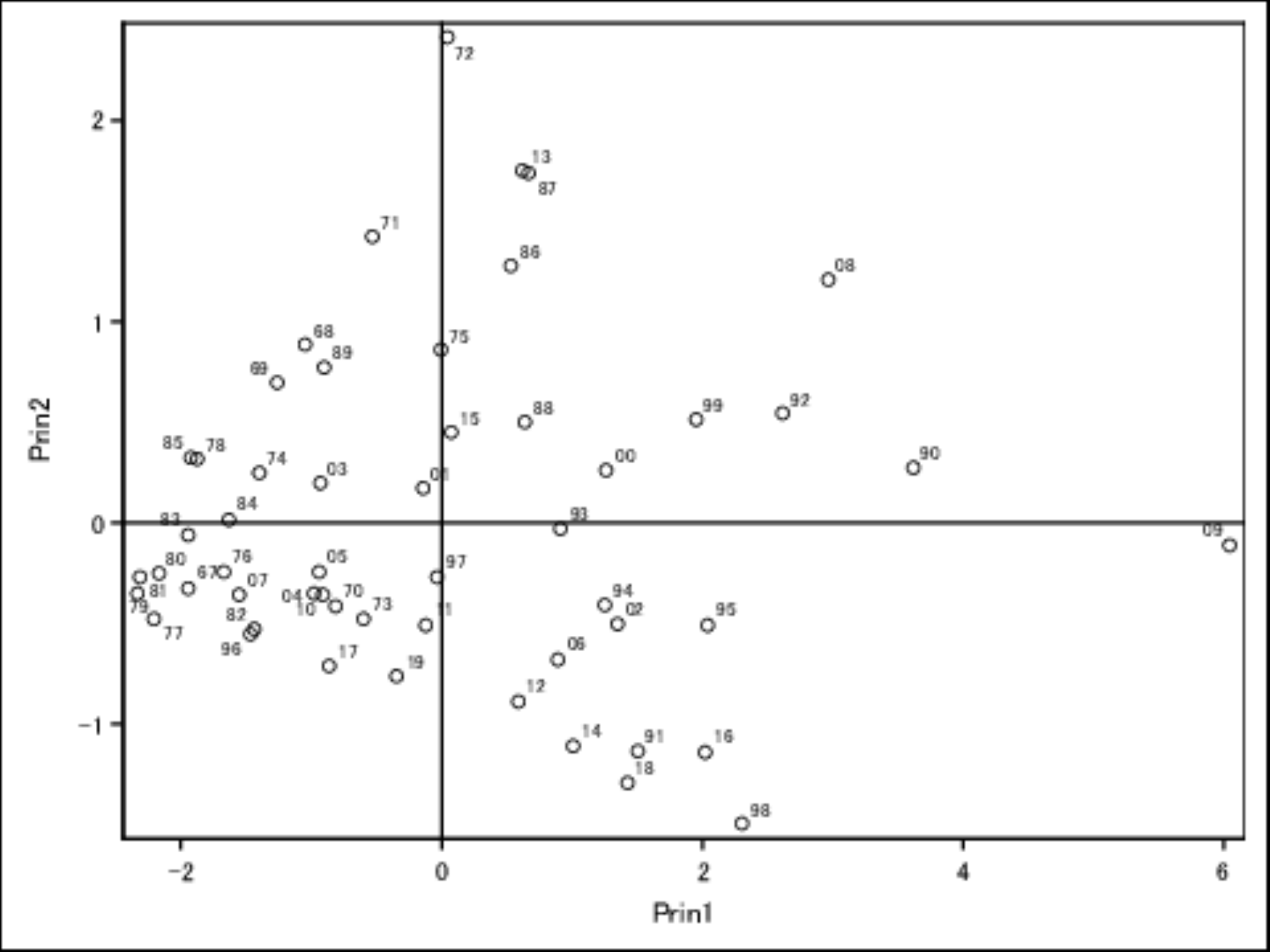}
\end{figure}

\begin{figure}
\caption{$\|V\|$ and $\|\nabla\|$}
\label{fig1}
\includegraphics[height=20cm,width=16cm]{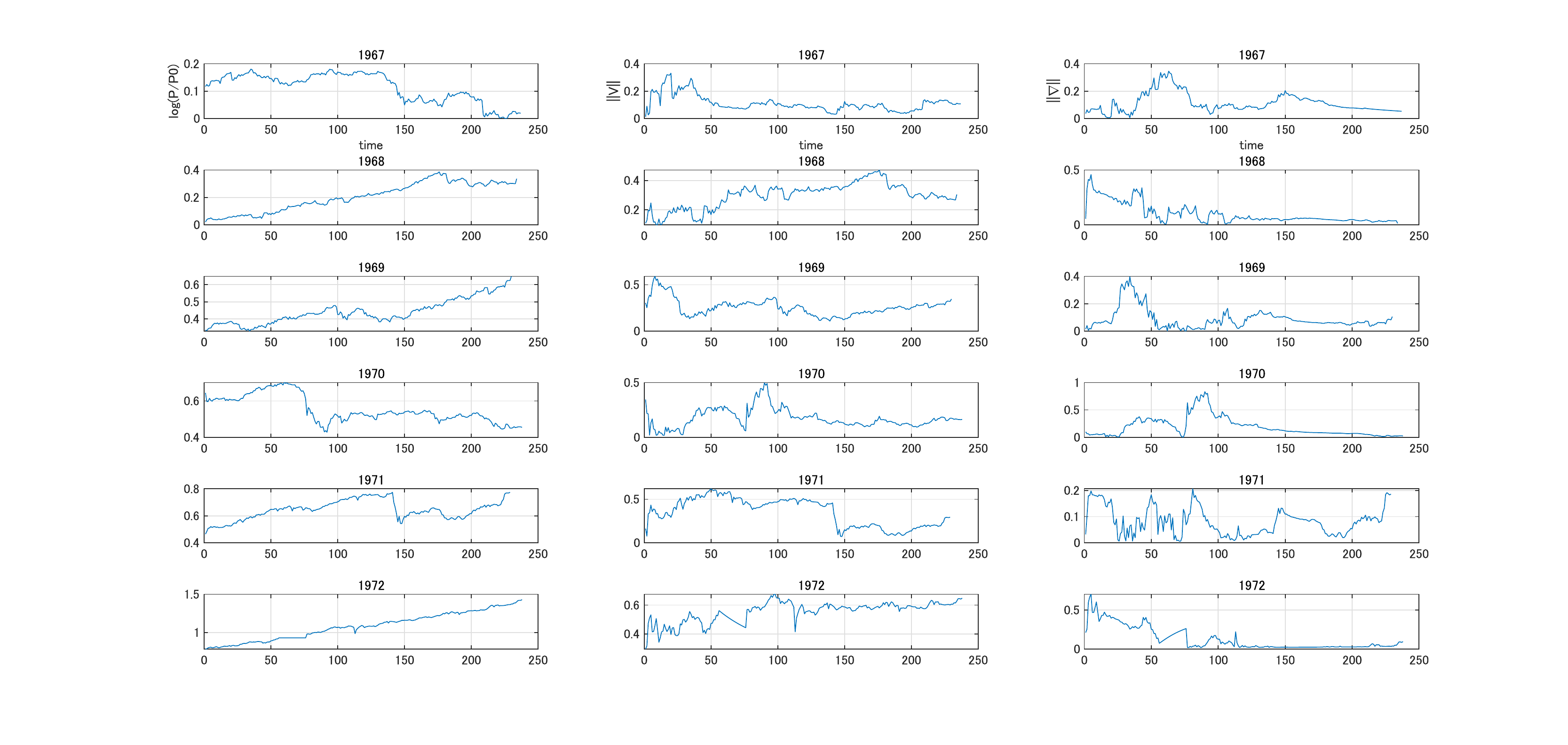}
\end{figure}

\begin{figure}
\caption{$\|V\|$ and $\|\nabla\|$}
\label{fig2}
\includegraphics[height=20cm,width=16cm]{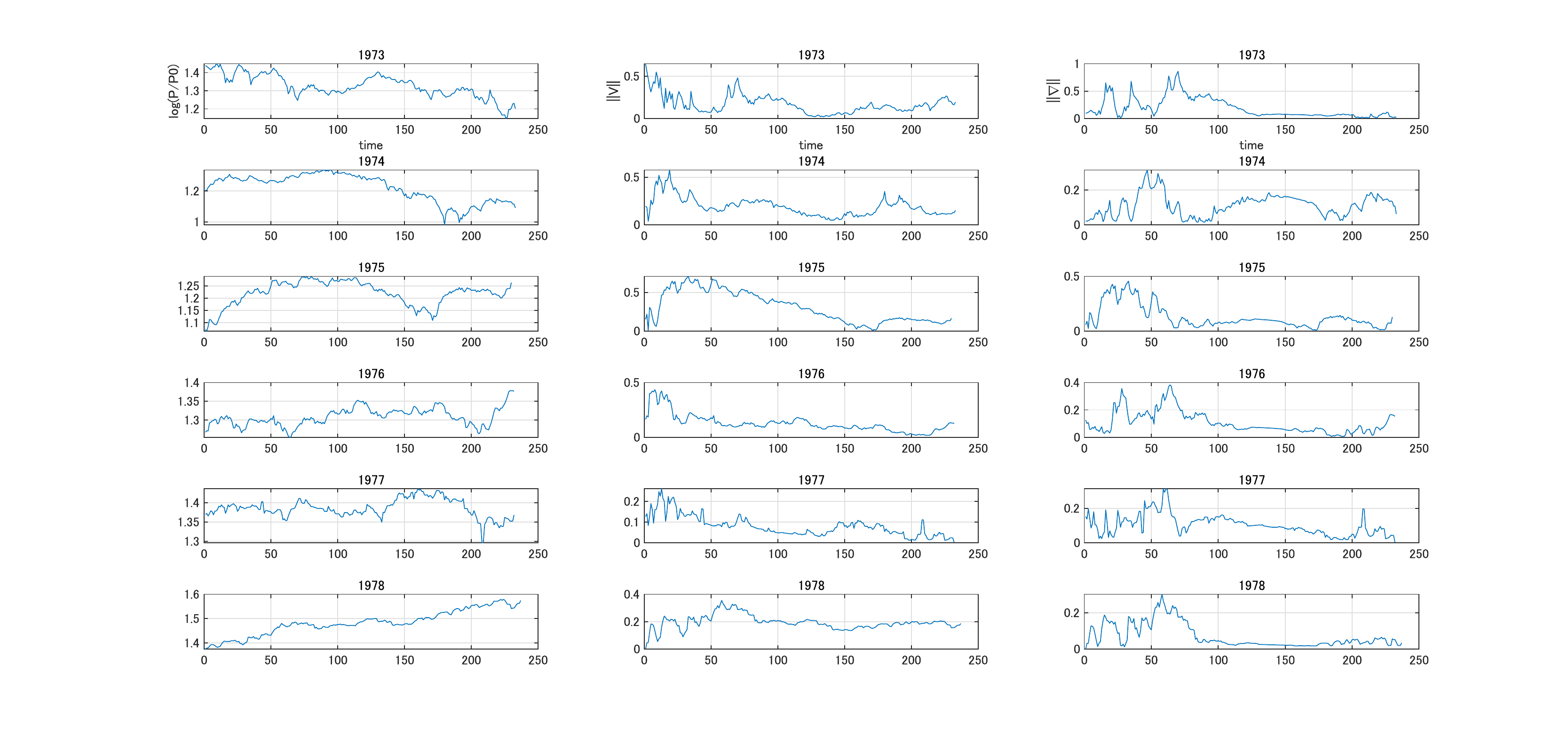}
\end{figure}

\begin{figure}
\caption{$\|V\|$ and $\|\nabla\|$}
\label{fig3}
\includegraphics[height=20cm,width=16cm]{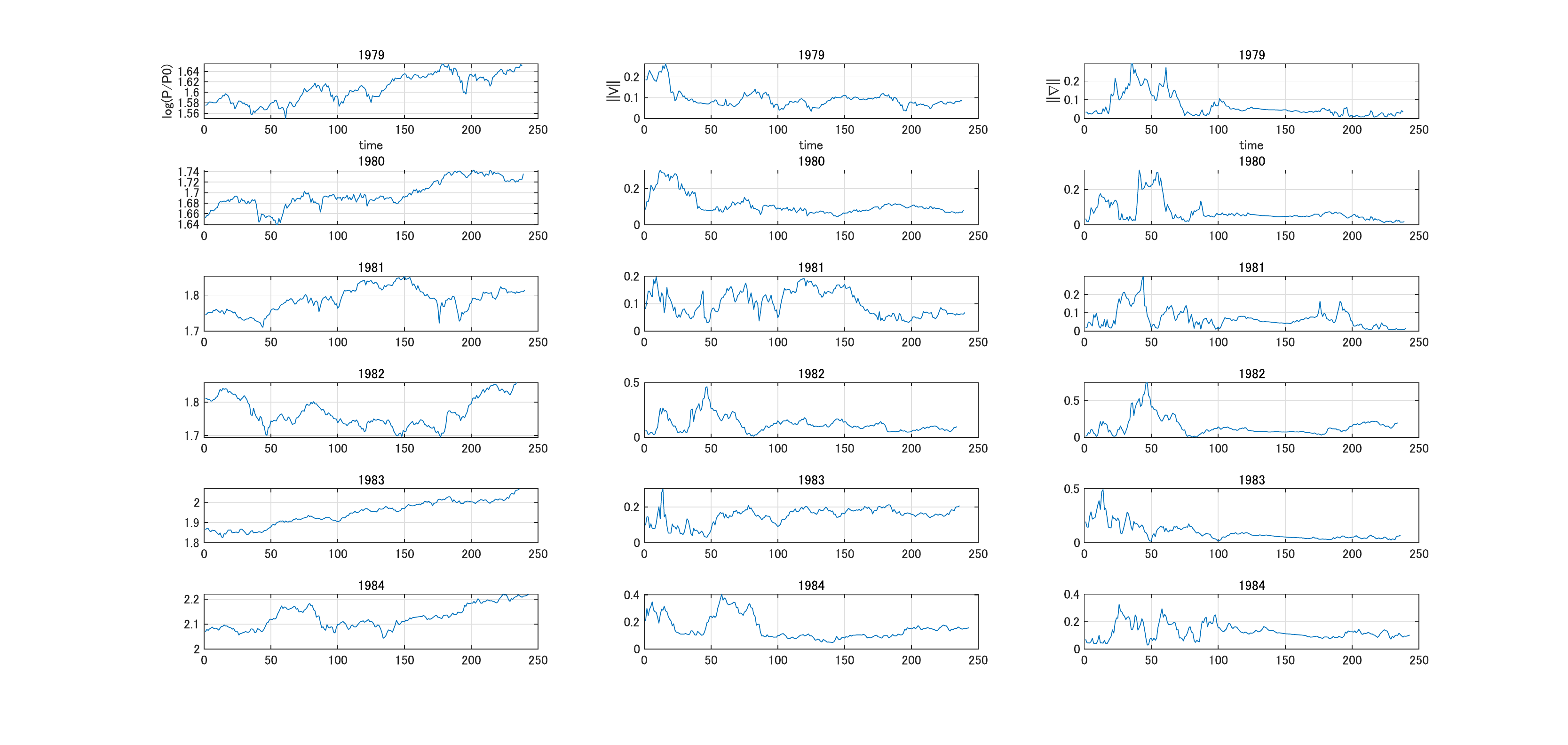}
\end{figure}

\begin{figure}
\caption{$\|V\|$ and $\|\nabla\|$}
\label{fig4}
\includegraphics[height=20cm,width=16cm]{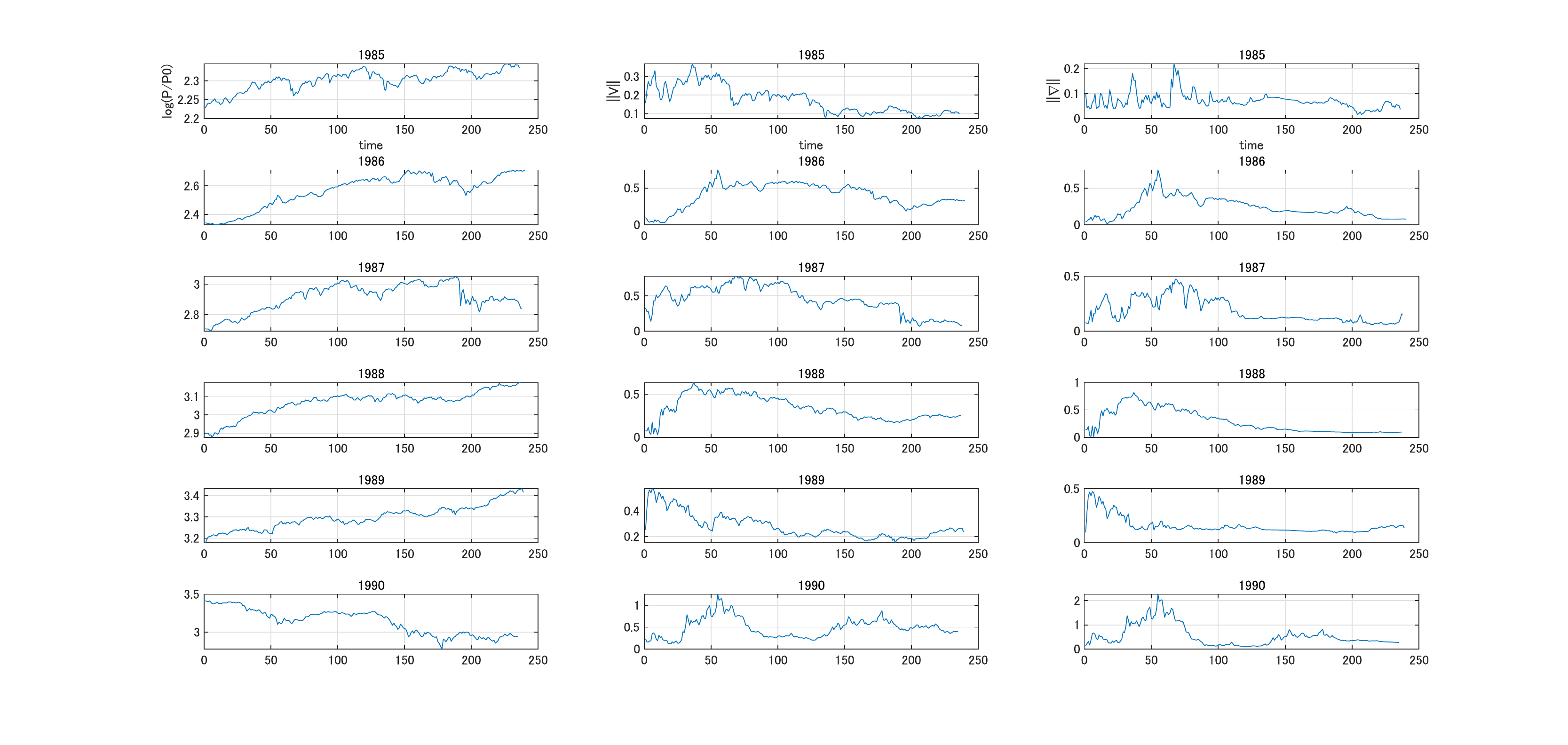}
\end{figure}

\begin{figure}
\caption{$\|V\|$ and $\|\nabla\|$}
\label{fig5}
\includegraphics[height=20cm,width=16cm]{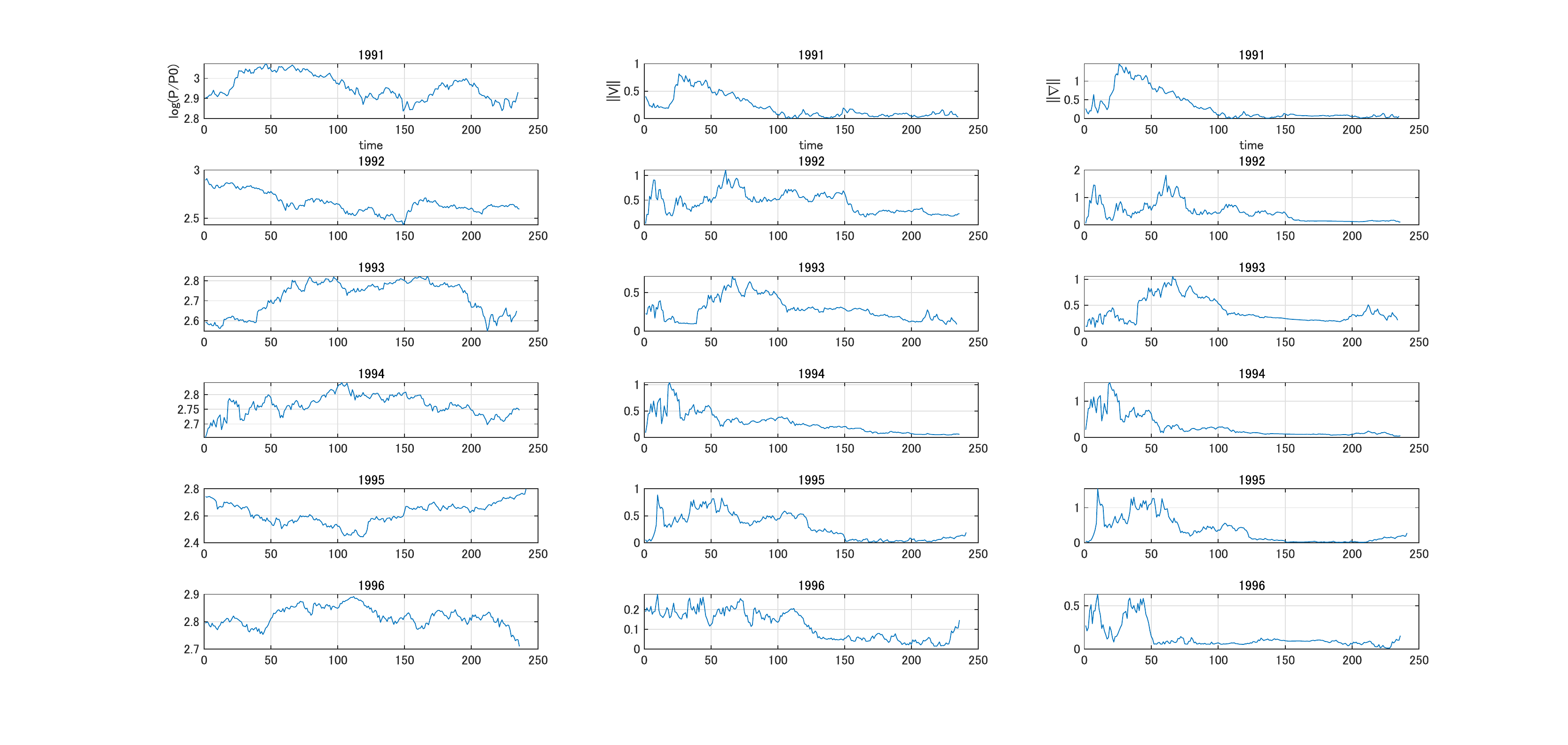}
\end{figure}

\begin{figure}
\caption{$\|V\|$ and $\|\nabla\|$}
\label{fig6}
\includegraphics[height=20cm,width=16cm]{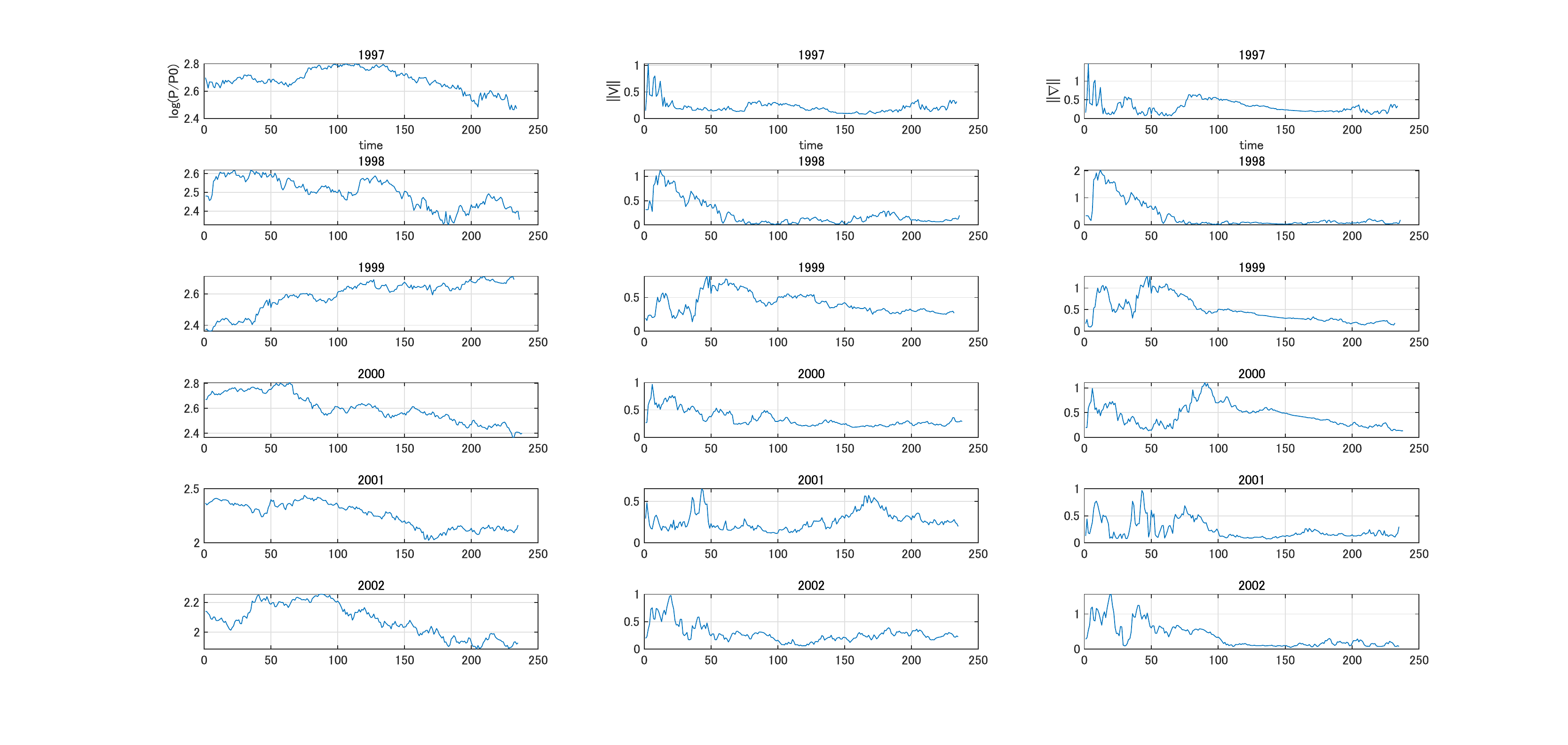}
\end{figure}

\begin{figure}
\caption{$\|V\|$ and $\|\nabla\|$}
\label{fig7}
\includegraphics[height=20cm,width=16cm]{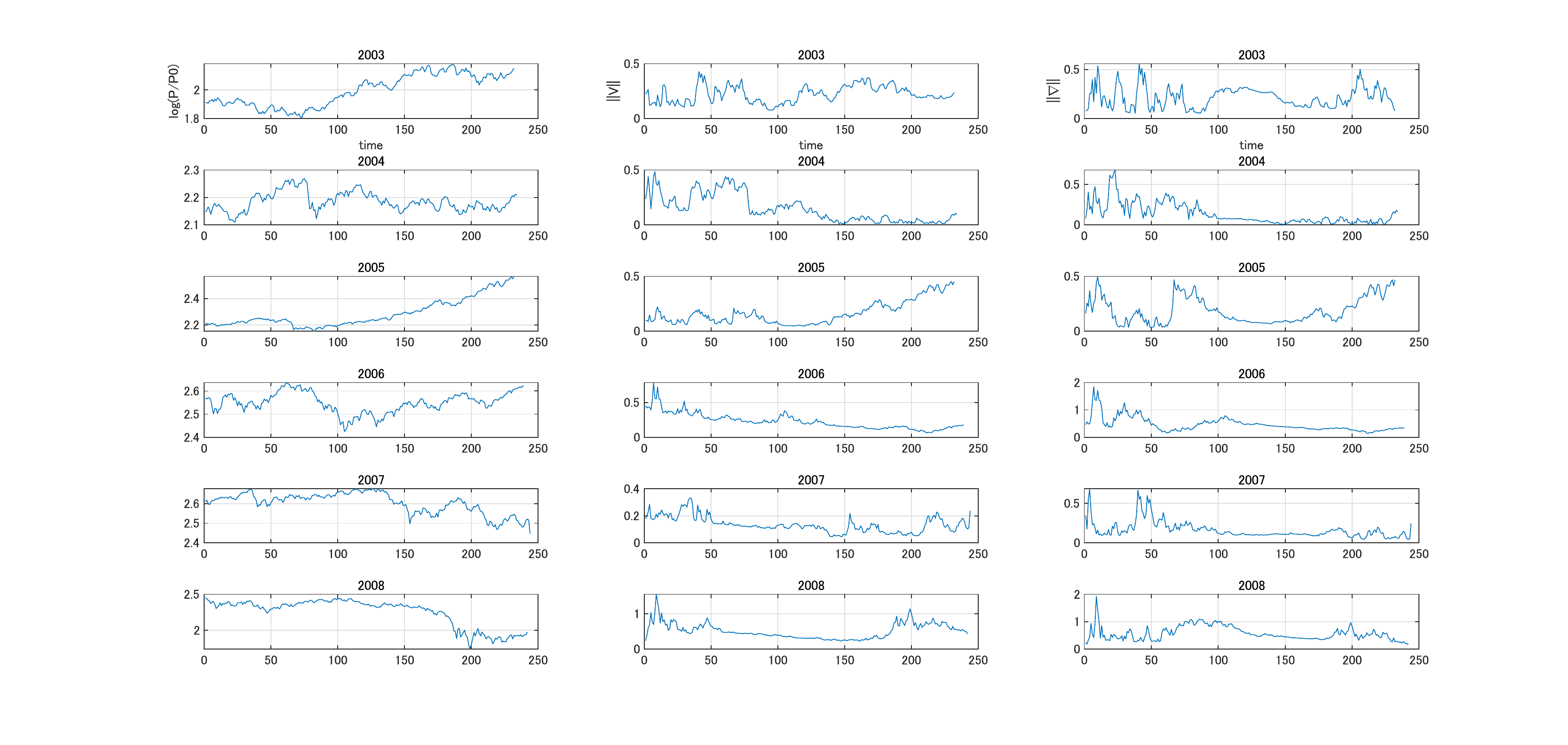}
\end{figure}

\begin{figure}
\caption{$\|V\|$ and $\|\nabla\|$}
\label{fig8}
\includegraphics[height=20cm,width=16cm]{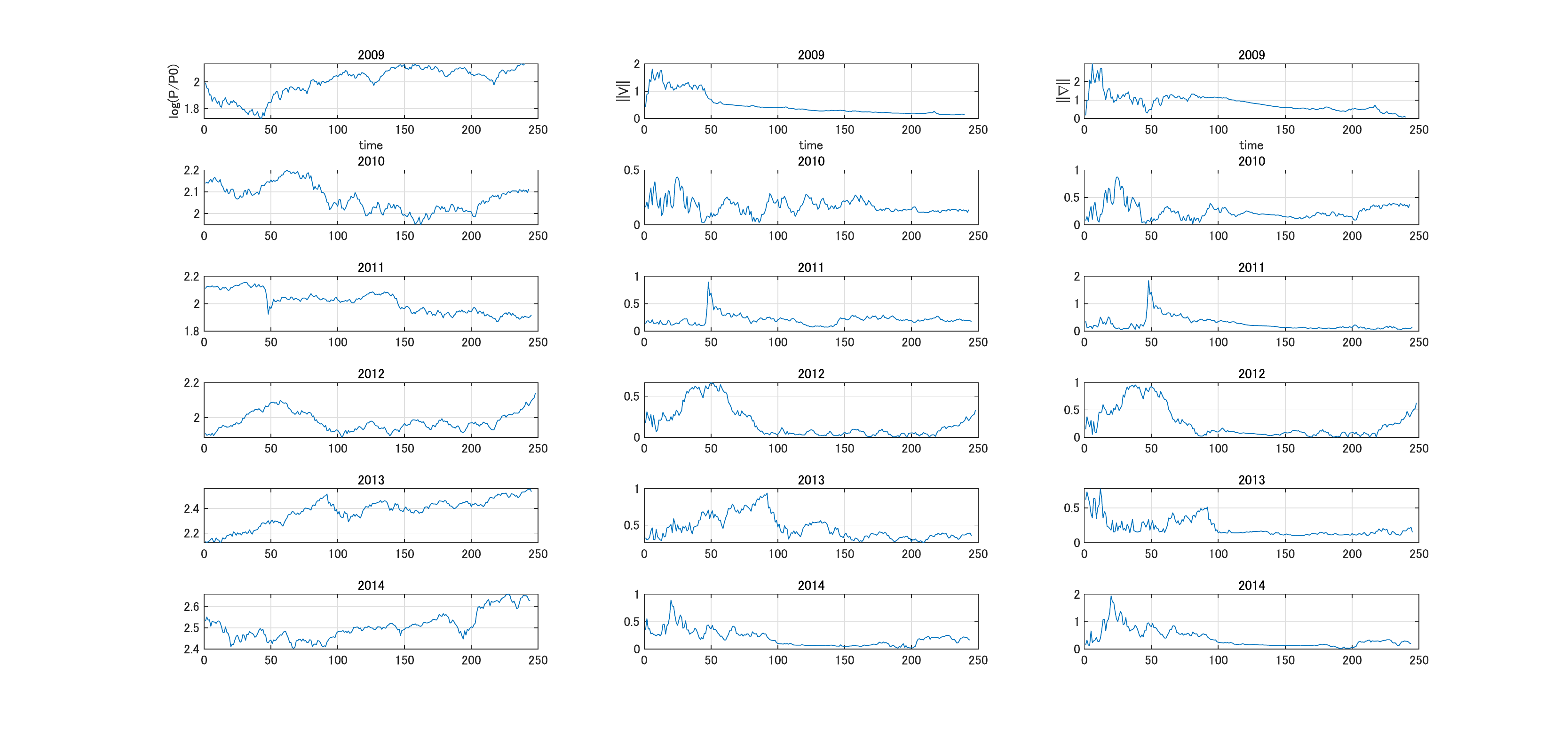}
\end{figure}

\begin{figure}
\caption{$\|V\|$ and $\|\nabla\|$}
\label{fig9}
\includegraphics[height=20cm,width=16cm]{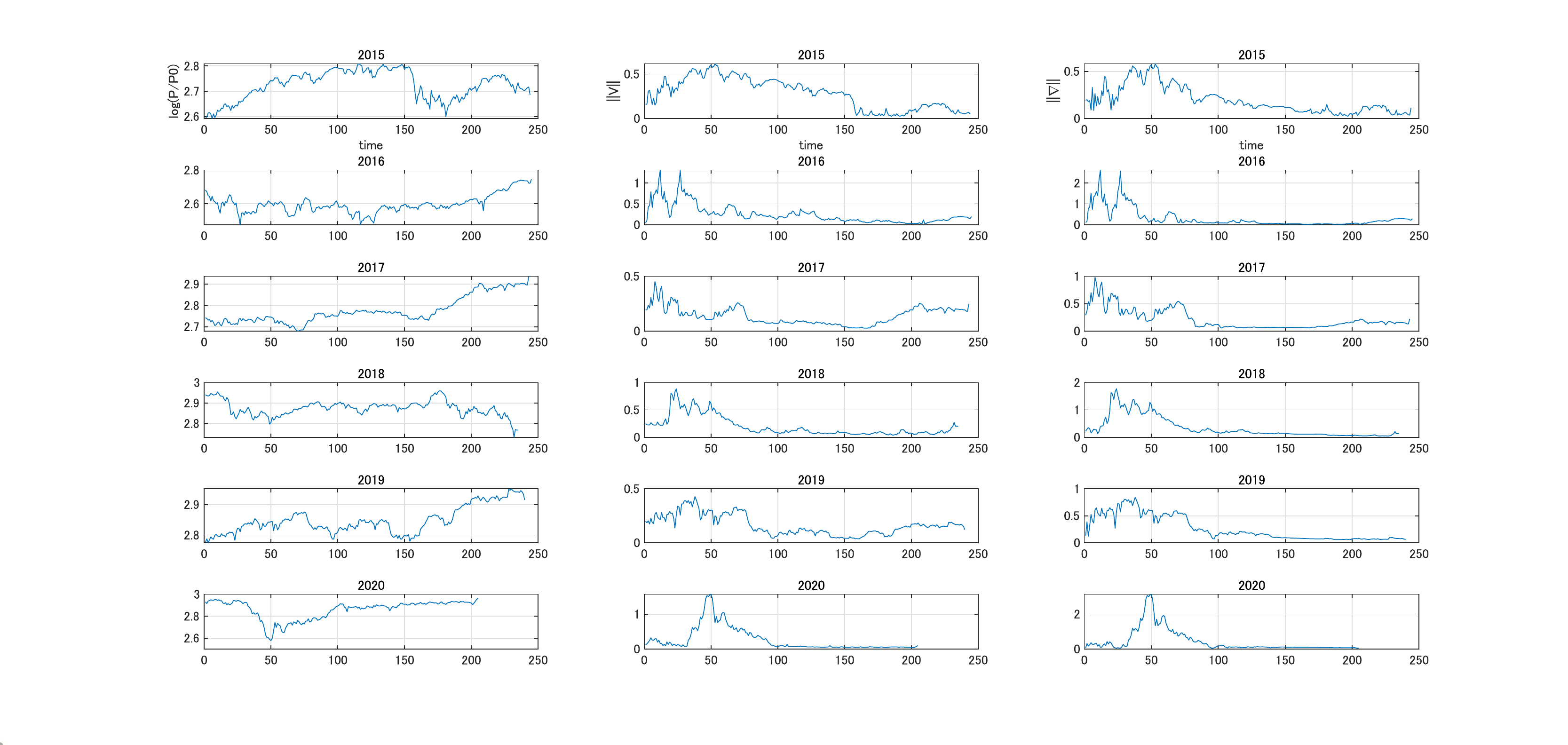}
\end{figure}

The mean of each times series can be analyzed by ANOVA, but it may be useful for investigating the time series properties in more detail to consider the simultaneous relation between the distribution of $\| V\|$ and that of $\|\nabla V\|$. To this end, for each year their distributions are simply explained by their sample means and standard deviations, and thereby the statistical property of each time series is characterized by the four factors; the two means and the two standard deviations. To analyze more than 50 set of time series, the principal component analysis is applied. The result is displayed in figure \ref{fig:prin}, where the horizontal axis stands for the first component and the vertical axis for the second component. The statistics tells us that the variation of more than 90 \% can be explained by the first and second components. 
According to the Hotelling's $t^2$ statistic with 4 degree of freedom, each member of $\{$1972(12.05),1989(10.23),1990(11.91),2008(17.15),2009(18.90)$\}$ is significant at the 5\% significant level (9.49), whose statistics are shown in parenthesis, suggesting that the time series of these years are quite different from the others. The unique feature of the crash observed in 2008 is reported by Beccar-Varela et al. (2017) in which the difference between the crashes in 2008 and 2010 are analyzed via the wavelet methodology. Our analysis also shows that $t^2$ statistic of 2010 (1.45) is quite different from that of 2008 (17.15). 

On the other hand, from an economical viewpoint, the Japanese economy had high gross domestic products in 1972. It attained the peak of the bubble economy in 1989 and subsequently had a hard crash. The sever crash was also experienced in 2008 caused by the Lehman's collapse. The significance of the statistics above suggests that these evidents are critical for the Japanese stock market.

%%%%%%%%%%%%%%%%%%%%%%%%%%%%%
\section{Conclusion}
We discussed  a geometric analysis of nonlinear dynamics and its application to financial time series. Nonlinear dynamics considered here are modeled by the second order ordinary differential equation. Since it is useful to geometrically analyze a curve, or a solution to the differential equation, in the differential phase space, the techniques in differential geometry is applied for the analysis of the dynamic behavior of the curve; the vector field on a manifold implied by the differential equation in the differential phase space and its covariant derivative.
We could derive information on the infinitesimal transition of the curve from the vector field, in other words the variability of the curve, and at the same time on the infinitesimal transition of the vector field, or the variability of the vector field, from the covariant derivative.

The analysis would work if the function form of the differential equation is specified beforehand. In real applications, however, we have almost no information on the specification especially in the case of more complex dynamics and thus the nonparametric method of filtering is used for estimation. 

We apply the above analysis to daily time series of the Japanese stock price index, called Nikkei225, sampled for more than fifty years in order to analyze the dynamic behavior of the index price from one year to another. Using the nonparametric estimation method, we estimate the vector fields and their covariant derivatives from the time series.
To compare the mean values of the vector fields and their covariant derivatives, the analysis of variance is applied, which revealed that the vector fields and the covariant derivatives catch different signals from the time series, but the two indicate that the drastic behavior occur from 2008 to 2009, corresponding to the years supposedly affected by the Lehman's collapse. We also conducted a principal component analysis to see the relation between the vector fields and the covariant derivative. As expected, the result implied that the covariant derivative can be used as the variability of the vector field.

%%%%%%%%%%%%%%%%%%%%%%%%%%%%%%%%%%%%%%%%%%%%%%%%%
\section{Appendix}
This appendix provides the detail formula of the nonparametric filtering. $\xi_k$ satisfies the following system and observation equations. The derivation can be seen in Shoji (2020) for example.
\begin{eqnarray}
\xi_{k+1}&=&F_k\xi_k+c_k+e_{k+1},\\
Y_{t_k}&=&H\xi_k+\varepsilon_{t_k}.
\end{eqnarray}
where,
\begin{eqnarray*}
F_k&=&I_8+G_kA_k,\\
c_k&=&G_kb_k,\\
e_{k+1}&=&G_k\epsilon_{k+1},
\end{eqnarray*}
with the $m\times m$ identity matrix $I_m$ and,
\begin{eqnarray*}
A_k&=&
\left(
\begin{array}{cccccc}
0& A_{1,2}& A_{1,3}& 0&\cdots& 0\\
0& A_{2,2}& A_{2,3}& 0&\cdots& 0\\
0&& \cdots& && 0\\
\vdots&&&&& \vdots\\
0&& \cdots& && 0
\end{array}
\right),\\
\left(
\begin{array}{cc}
A_{1,2}& A_{1,3}\\
A_{2,2}& A_{2,3}\\
\end{array}
\right)&=&
(J_{t_k})^{-1}\left\{\exp( J_{t_k}\Delta t)-I_2\right\},\\
J_{t_k}&=&
\left(
\begin{array}{cc}
0& 1\\
Y_{t_k|t_k}^{(1,0)}& Y_{t_k|t_k}^{(0,1)}
\end{array}
\right),\\
G_k&=&
\left(
\begin{array}{cccccc}
1& 0& 0& \cdots&& 0\\
0& 1& 0& \cdots&& 0\\
Y_{t_k|t_k}^{(1,0)}& Y_{t_k|t_k}^{(0,1)}& 1& \cdots&& 0\\
\vdots& \vdots& \vdots& \ddots&& \vdots\\
\theta_{1}& \theta_2& 0& &\ddots& 0\\
\theta_{2}& \theta_3& 0& &\cdots& 1
\end{array}
\right),\\
b_k&=&
\left(
\begin{array}{c}
(J_{t_k}^{-1})^2\left\{\exp(J_{t_k}\Delta t)-I_2-J_{t_k}\Delta t\right\}
M_{t_k}\\
({\sigma_1^2\over 2}Y_{t_k|t_k}^{(2,0)}+{\sigma_2^2\over 2}Y_{t_k|t_k}^{(0,2)})\Delta t\\
\vdots\\
0\\
\end{array}
\right),\\
e_{k+1}&=&
\left(
\begin{array}{c}
\int_{t_k}^{t_{k+1}}\exp\{J_{t_k}(t_{k+1}-u)\}S dB_u\\
0\\
\vdots\\
0\\
\end{array}
\right),\\
M_{t_k}&=&
\left(
\begin{array}{c}
0\\
{\sigma_1^2\over 2}Y_{t_k|t_k}^{(2,0)}+{\sigma_2^2\over 2}Y_{t_k|t_k}^{(0,2)}
\end{array}
\right),\ 
S=
\left(
\begin{array}{c}
\sigma_1\\
\sigma_2\\
\end{array}
\right),\\
H&=&
\left(
\begin{array}{cccc}
1& 0& \cdots& 0\\
\end{array}
\right).
\end{eqnarray*}

%%%%%%%%%%%%%%%%%%%%%%%%%%%%%%%%%%%%%%%%%%%%%%%%%

%\renewcommand{\labelitemi}{\ }
%\begin{itemize}

%\end{itemize}

%%%%%%%%%%%%%%%%%%%%%%%%%%%%%%%%%%%%%%%%%%%%%%%%%
% TABLE
%%%%%%%%%%%%%%%%%%%%%%%%%%%%%%%%%%%%%%%%%%%%%%%%%
\clearpage
\begin{table}
\caption{Parameter estimates of the state space model: the parameters are estimated from daily time series from 1965 to 2020.}
\label{tbl:param}
\begin{tabular}{lllllll}
\hline
$\sigma_1$& $\sigma_2$& $\sigma_\varepsilon$& $\theta_0$& $\theta_1$& $\theta_2$& $\theta_3$\\ 
\hline

0.202696& 0.083083& 0.001351& 0.022526& 0.012242& 0.002956& 0.000934\\
\hline
\end{tabular}
\end{table}

\bigskip
%\clearpage
\begin{longtable}{llllllll}
\caption{Summary statistics: columns of $p$, $\|V\|$, and $\|\nabla V\|$ stand for their sample means and standard deviations in parenthesis.\label{tbl:sumstat}}
\\
\hline\hline
Year& $N$& ${p=\log(P/P_0)}$& ${\|V\|}$& ${\|\nabla V\|}$& $\rho_{p,\|V\|}$& $\rho_{p,\|\nabla V\|}$& $\rho_{\|V\|,\|\nabla V\|}$\\
\hline
\endfirsthead
\hline
Year& $N$& ${p=\log(P/P_0)}$& ${\|V\|}$& ${\|\nabla V\|}$& $\rho_{p,\|V\|}$& $\rho_{p,\|\nabla V\|}$& $\rho_{\|V\|,\|\nabla V\|}$\\
\hline
\endhead
\input{figtbl.tex}
\end{longtable}
%\end{table}
%%%%%%%%%%%%%%%%%%%%%%%%%%%%%%%%%%%%%%%%%%%%%%%%%
% FIGURE
%%%%%%%%%%%%%%%%%%%%%%%%%%%%%%%%%%%%%%%%%%%%%%%%%
\clearpage
%\begin{landscape}
%\end{landscape}
%%%%%%%%%%%%%%%%%%%%%%%%%%%%

%\end{document}

\clearpage
%\begin{landscape}
%\end{landscape}
%%%%%%%%%%%%%%%%%%%%%%%%%%%%
\clearpage
%%%%%%%%%%%%%%%%%%%%%%%%%%%%
\clearpage
%%%%%%%%%%%%%%%%%%%%%%%%%%%%
\clearpage
%%%%%%%%%%%%%%%%%%%%%%%%%%%%
\clearpage
%%%%%%%%%%%%%%%%%%%%%%%%%%%%
\clearpage
%%%%%%%%%%%%%%%%%%%%%%%%%%%%
\clearpage
%%%%%%%%%%%%%%%%%%%%%%%%%%%%
\clearpage
%%%%%%%%%%%%%%%%%%%%%%%%%%%%
\clearpage

%%%%%%%%%%%%%%%%%%%%%%%%%%%%%%%%%%%%%%%%%%%%%%%%%

\end{document}

%% file: figtbl.tex
1967& 237& 0.11394& 0.15603& 0.19885& 0.79246& -0.76849& -0.28617\\
&& (0.05282)& (0.03462)& (0.04894)\\
1968& 234& 0.20401& 0.13017& 0.09497& 0.93051& 0.14746& 0.46238\\
&& (0.10449)& (0.04051)& (0.06230)\\
1969& 230& 0.44161& 0.16810& 0.07584& 0.78851& 0.62869& 0.89490\\
&& (0.07174)& (0.02060)& (0.04384)\\
1970& 238& 0.55242& 0.16726& 0.17174& 0.81608& -0.65072& -0.16811\\
&& (0.07241)& (0.03025)& (0.05996)\\
1971& 229& 0.64372& 0.15215& 0.10907& 0.77822& -0.08629& 0.24937\\
&& (0.07090)& (0.02065)& (0.06177)\\
1972& 238& 1.08309& 0.25833& 0.27824& 0.95428& 0.87077& 0.97758\\
&& (0.17850)& (0.03767)& (0.05803)\\
1973& 233& 1.32721& 0.20420& 0.14913& 0.62832& -0.08809& 0.65667\\
&& (0.06350)& (0.04337)& (0.10775)\\
1974& 233& 1.22120& 0.18349& 0.23734& -0.86304& -0.96110& 0.94109\\
&& (0.08899)& (0.03575)& (0.13871)\\
1975& 230& 1.21959& 0.14439& 0.11605& -0.73858& -0.96215& 0.75300\\
&& (0.04944)& (0.01524)& (0.09246)\\
1976& 232& 1.30884& 0.13046& 0.05042& 0.01241& -0.45551& -0.37119\\
&& (0.02415)& (0.00468)& (0.03967)\\
1977& 232& 1.38560& 0.12527& 0.05803& -0.30551& -0.77135& 0.16542\\
&& (0.02370)& (0.00614)& (0.05257)\\
1978& 237& 1.48319& 0.12163& 0.03382& 0.62994& -0.24984& 0.18655\\
&& (0.05094)& (0.00567)& (0.02298)\\
1979& 238& 1.60694& 0.11488& 0.02531& -0.30958& -0.18174& 0.21529\\
&& (0.02597)& (0.00633)& (0.01635)\\
1980& 239& 1.69731& 0.10669& 0.02629& -0.16300& -0.47961& -0.27794\\
&& (0.02589)& (0.00476)& (0.01567)\\
1981& 240& 1.78806& 0.10333& 0.04311& 0.33302& -0.10153& 0.15160\\
&& (0.03344)& (0.00773)& (0.03131)\\
1982& 234& 1.76935& 0.10525& 0.10958& -0.63420& -0.92411& 0.82596\\
&& (0.04251)& (0.01139)& (0.06062)\\
1983& 236& 1.94511& 0.11324& 0.07521& 0.55910& 0.70720& 0.97579\\
&& (0.06043)& (0.01102)& (0.03005)\\
1984& 243& 2.12752& 0.11085& 0.08725& 0.32730& 0.37692& 0.98446\\
&& (0.04605)& (0.02767)& (0.05379)\\
1985& 236& 2.30233& 0.10388& 0.09242& -0.43608& -0.38511& 0.99733\\
&& (0.02639)& (0.02717)& (0.04666)\\
1986& 240& 2.56674& 0.16419& 0.19805& 0.62706& 0.68673& 0.99472\\
&& (0.11533)& (0.05883)& (0.08892)\\
1987& 238& 2.91420& 0.19924& 0.25508& 0.35484& 0.36332& 0.99961\\
&& (0.09050)& (0.08360)& (0.10496)\\
1988& 237& 3.06892& 0.14109& 0.18588& 0.62382& 0.62999& 0.99984\\
&& (0.06763)& (0.03660)& (0.04599)\\
1989& 239& 3.29927& 0.18587& 0.23745& 0.18610& 0.10810& 0.99674\\
&& (0.05444)& (0.01935)& (0.02354)\\
1990& 235& 3.13780& 0.17219& 0.17335& -0.74869& -0.64851& 0.97746\\
&& (0.16557)& (0.08576)& (0.11722)\\
1991& 235& 2.95889& 0.10122& 0.09959& -0.76625& -0.83886& 0.99131\\
&& (0.06223)& (0.05429)& (0.07596)\\
1992& 236& 2.65733& 0.20611& 0.28397& -0.55113& -0.62896& 0.99516\\
&& (0.10379)& (0.09120)& (0.13054)\\
1993& 234& 2.71791& 0.08241& 0.10836& -0.45433& -0.56247& 0.98887\\
&& (0.08164)& (0.03830)& (0.05935)\\
1994& 236& 2.76409& 0.04059& 0.04949& 0.49599& 0.38832& 0.98249\\
&& (0.03508)& (0.02214)& (0.02975)\\
1995& 241& 2.61894& 0.10679& 0.14648& -0.77843& -0.79592& 0.99564\\
&& (0.07858)& (0.07076)& (0.11311)\\
1996& 236& 2.81897& 0.08597& 0.10841& 0.52135& 0.53019& 0.99930\\
&& (0.03476)& (0.04921)& (0.06302)\\
1997& 234& 2.67966& 0.09648& 0.13592& -0.94590& -0.95634& 0.99916\\
&& (0.08169)& (0.06227)& (0.09371)\\
1998& 236& 2.49937& 0.13432& 0.20371& -0.80466& -0.82230& 0.99929\\
&& (0.07523)& (0.05554)& (0.08597)\\
1999& 232& 2.59104& 0.09451& 0.11055& -0.12144& -0.32972& 0.95776\\
&& (0.09204)& (0.02909)& (0.05429)\\
2000& 238& 2.60442& 0.11250& 0.15884& -0.35775& -0.53800& 0.97795\\
&& (0.11250)& (0.04828)& (0.07466)\\
2001& 235& 2.25475& 0.22490& 0.35998& -0.76244& -0.79816& 0.99810\\
&& (0.11762)& (0.07296)& (0.12222)\\
2002& 235& 2.07913& 0.17321& 0.27802& -0.91618& -0.92610& 0.99588\\
&& (0.10896)& (0.07273)& (0.14253)\\
2003& 232& 1.99581& 0.14762& 0.19384& -0.70269& -0.79458& 0.98231\\
&& (0.11066)& (0.05674)& (0.14507)\\
2004& 234& 2.18481& 0.09315& 0.07110& 0.65998& 0.57862& 0.92957\\
&& (0.03444)& (0.03006)& (0.05254)\\
2005& 232& 2.28474& 0.10774& 0.09896& 0.97701& 0.93551& 0.98000\\
&& (0.10598)& (0.05761)& (0.09767)\\
2006& 239& 2.54942& 0.12557& 0.15200& 0.60467& 0.62754& 0.99847\\
&& (0.04355)& (0.07000)& (0.10034)\\
2007& 244& 2.60200& 0.07522& 0.09735& 0.11316& -0.08257& 0.97474\\
&& (0.05391)& (0.03247)& (0.04664)\\
2008& 242& 2.24833& 0.24285& 0.39196& -0.95172& -0.96043& 0.99918\\
&& (0.19384)& (0.15961)& (0.27381)\\
2009& 240& 1.99972& 0.17031& 0.25348& -0.94603& -0.96866& 0.99483\\
&& (0.10887)& (0.14685)& (0.28070)\\
2010& 243& 2.07242& 0.09155& 0.10721& -0.58194& -0.75814& 0.93580\\
&& (0.06164)& (0.02570)& (0.07378)\\
2011& 245& 2.01027& 0.10867& 0.15349& -0.89222& -0.94808& 0.96974\\
&& (0.07808)& (0.03665)& (0.09196)\\
2012& 248& 1.97875& 0.09179& 0.10652& -0.30963& -0.66663& 0.88481\\
&& (0.05352)& (0.02108)& (0.06260)\\
2013& 245& 2.37491& 0.20546& 0.27258& 0.23766& 0.27363& 0.99886\\
&& (0.10935)& (0.05528)& (0.08079)\\
2014& 244& 2.50734& 0.09156& 0.11397& 0.82050& 0.83752& 0.99814\\
&& (0.06346)& (0.04354)& (0.06489)\\
2015& 244& 2.72446& 0.12222& 0.16575& 0.53827& 0.55264& 0.99936\\
&& (0.05637)& (0.06530)& (0.08873)\\
2016& 245& 2.59804& 0.06888& 0.09501& -0.20951& -0.23297& 0.99905\\
&& (0.05373)& (0.04408)& (0.06191)\\
2017& 243& 2.77571& 0.09781& 0.13567& 0.88269& 0.88282& 0.99988\\
&& (0.06262)& (0.03926)& (0.05229)\\
2018& 235& 2.87413& 0.08050& 0.10664& 0.69871& 0.73871& 0.98092\\
&& (0.03741)& (0.03765)& (0.05624)\\
2019& 240& 2.84790& 0.05391& 0.05684& 0.74391& 0.85238& 0.96355\\
&& (0.04462)& (0.02534)& (0.04565)\\
\hline

%% file: geomet20Dec21.bbl
\begin{thebibliography}{99}
%[EKF]
\bibitem{and} Anderson B. D.O., Moor J. B.:
Optimal Filtering.
Prentice-Hall, New Jersey (1979)

%[Finance]
\bibitem{bec} Beccar-Varela, M. P. et al.:
Analysis of the Lehman Brothers collapse and the Flash Crash event by applying wavelets methodologies.
Physica A 474, 162-171 (2017)

%[Finance]
\bibitem{chia} Chiarella, C., Hung, H, To, T. D.:
The volatility structure of the fixed income market under the HJM framework: A nonlinear filtering approach.
Computational Statistics \& Data Analysis 53, (2009)

%[Finance]
\bibitem{date} Date, P., Ponomareva, K.:
Linear and non-linear filtering in mathematical finance: a review.
IMA Journal of Management Mathematics 22, 195- 211 (2011)

%[LLA]
\bibitem{fan:96} Fan, J., Gijbels, I.:
Local Polynomial Modeling and Its Applications. 
Chapman and Hall, London.  (1996) 

%[LLA]
\bibitem{fan:05} Fan, J., Yao, Q. W.:
Nonlinear Time Series: Nonparametric and Parametric Methods.
Springer.  (2005) 

%[LLA]
\bibitem{fan:03} Fan, J., Zhang, C.:
A reexamination of diffusion estimators with applications to financial model validation.
Journal of the American Statistical Association {\bf 98}, 118-134 (2003)

%[Card]
\bibitem{gru} Grudzinski, K. and Zebrowski, J. J.: 
Modeling cardiac pacemakers with relaxation oscillators. 
Physica A 336, 153-162 (2004)

%[Epid]
\bibitem{gao} Gao, N. et al.:
Dynamics of a stochastic SIS epidemic model with nonlinear incidence rates
Advances in Difference Equations 2019:41, (2019)

%[Card]
\bibitem{goi} Gois, S.F.S.M. and Savi, M. A.:
An analysis of heart rhythm dynamics using a three-coupled oscillator model.
Chaos, Solitons and Fractals 41, 2553-2565 (2009)

%[LLF]
\bibitem{havli} Havlicek, M., Friston, K.J., et al.:
Dynamic modeling of neuronal responses in fMRI using cubature Kalman filtering.
Neuroimage 56, 2109-2128 (2011)

%[EKF]
\bibitem{jaz} Jazwinski A.H.:
Stochastic Processes and Filtering Theory.
Academic Press, New York (1970)

%[LL]
\bibitem{jime} Jimenez, J.C., Shoji, I., Ozaki, T.:
Simulation of stochastic differential equations through the local linearization method. A comparative study.
Journal of Statistical Physics 94, 587-602 (1999)

%[Astro]
\bibitem{mis} Misra, R., Zdziarski, A.A.;
Damped harmonic oscillator interpretation of the soft-state power spectra of Cyg X-1.
MNRAS. 387, 915?920 (2008)

%[LL,LLF]
\bibitem{ozaki} Ozaki, T.:
Time series modeling of neuroscience data.
CRC Press (2012)

%[Astro]
\bibitem{phil} Phillipson R. A., Boyd P. T., Smale A.P.:
The chaotic long-term X-ray variability of 4U 1705-44.
MNRAS. 477, 5220-5237 (2018)

%[LLF]
\bibitem{riera} Riera, J.J.,  Watanabe, J. et al.:
A state-space model of the hemodynamic approach: nonlinear filtering of BOLD signals.
NeuroImage 21, 547-567 (2004)

\bibitem{shoji:98} Shoji, I.:
A comparative study of maximum likelihood estimators for nonlinear dynamical system models.
International Journal of Control 71, 391-404 (1998)

%[NPF]
\bibitem{shoji:2000} Shoji, I.:
A nonparametric method of estimating nonlinear dynamical system models.
Physics Letters A 277, 159-168 (2000)

\bibitem{shoji:2020} Shoji, I.:
Nonparametric filtering for stochastic nonlinear oscillations,
forthcoming in Physical Review E (2020)

\bibitem{stm:2020} Shoji, I., Takata, T., and Mizumoto, Y.
A geometric method of analysis for the light curves of active galactic nuclei,
Monthly Notices of the Royal Astronomical Society, 495, 338-349 (2020).

%[LL,Neuro]
\bibitem{ste} Stephan, K.E., Kasper, L., et al.:
Nonlinear dynamic causal models for fMRI.
Neuroimage 42, 649-662 (2008)

%[LLF]
\bibitem{valdes} Valdes-Sosa, P.A., Sanchez-Bornot, J.M., et al.:
Model driven EEG/fMRI fusion of brain oscillations.
Human Brain Mapping 30, 2701-2721 (2009)

%[Epid]
\bibitem{wan} Wang, W., et al.:
A stochastic differential equation SIS epidemic model incorporating Ornstein?Uhlenbeck process.
Physica A 509, 921-936 (2018)

%[Card]
\bibitem{zeb} Zebrowski, J. J., et al.: 
Nonlinear oscillator model reproducing various phenomena in the dynamics of the conduction system of the heart.
Chaos 17, 015121-11 (2007)

%[UKF,Neuro]
\bibitem{che} Che, Y, Liu, B., et al.:
Robust stabilization control of bifurcations in Hodgkin-Huxley model with aid of unscented Kalman filter.
Chaos, Solitons \& Fractals 101, 92-99 (2017)

%[UKF]
\bibitem{mar} Mariani, S., Ghisi, A.:
Unscented Kalman filtering for nonlinear structural dynamics.
Nonlinear Dynamics 49, 131-150 (2007)

%[UKF,Bio]
\bibitem{qua} Quach, M., Brunel, N., d'Alche-Buc, F.:
Estimating parameters and hidden variables in non-linear state-space models based on ODEs for biological networks inference.
Bioinformatics 23, 3209-3216 (2007)

%[UKF]
\bibitem{sit} Sitz, A., Schwarz, U., et al.:
Estimation of parameters and unobserved components for nonlinear systems from noisy time series.
Physical Review E 66, 016210 (2002)

\bibitem{chen} Chen, Z.; Bayesian Filtering: From Kalman Filters to Particle Filters, and Beyond. Statistics 182, 1-69 (2003)

%%%%%%%%%%%%%%%%%%%%%%%%%%%%%%%%%%%%%%%%%%%%%%%%%

\end{thebibliography}
